\documentclass[prc,aps,floats,showpacs,twocolumn,twoside,preprintnumbers,
superscriptaddress,floatfix]{revtex4}
\usepackage{graphicx,epsfig,rotating}
\newcommand{\gsim}{\stackrel{\scriptstyle >}{\phantom{}_{\sim}}}
\newcommand{\lsim}{\stackrel{\scriptstyle <}{\phantom{}_{\sim}}}
\newcommand{\turnt}[1]{\begin{turn}{90}#1\end{turn}}
\begin{document}
%\begin{frontmatter}
%v12 of jan.23
%corrections suggested by tuebingen - jan 24., done
%corrections suggested by dima - jan 24, not done
\title{Constraints on the high-density  nuclear equation of state
from the phenomenology of
compact stars and heavy-ion collisions }
\date{\today}
\author{T.~Kl\"ahn}
\email{thomas.klaehn@uni-rostock.de}
\affiliation{
Institut f\"ur Physik,
Universit\"at Rostock,
18051 Rostock,
Germany}
\affiliation{
Institut f\"ur Theoretische Physik,
Universit\"at T\"ubingen,\\
72076 T\"ubingen,
Germany}
\author{D.~Blaschke}
\email{blaschke@theory.gsi.de}
\affiliation{
Gesellschaft f\"ur Schwerionenforschung mbH (GSI),
64291 Darmstadt,
Germany}
\affiliation{
Bogoliubov  Laboratory of Theoretical Physics,\\
Joint Institute for Nuclear Research,
141980 Dubna,
Russia}
\author{S.~Typel}
\affiliation{
Gesellschaft f\"ur Schwerionenforschung mbH (GSI),
64291 Darmstadt,
Germany}
\author{E.N.E.~van~Dalen}
\affiliation{
Institut f\"ur Theoretische Physik,
Universit\"at T\"ubingen,\\
72076 T\"ubingen,
Germany}
\author{A.~Faessler}
\affiliation{
Institut f\"ur Theoretische Physik,
Universit\"at T\"ubingen,\\
72076 T\"ubingen,
Germany}
\author{C.~Fuchs}
\affiliation{
Institut f\"ur Theoretische Physik,
Universit\"at T\"ubingen,\\
72076 T\"ubingen,
Germany}
\author{T.~Gaitanos}
\affiliation{
Department f\"ur Physik,
Universit\"at M\"unchen,
85748 Garching,
Germany}
\author{H.~Grigorian}
\affiliation{
Institut f\"ur Physik,
Universit\"at Rostock,
18051 Rostock,
Germany}
\affiliation{
Department of Physics,
Yerevan State University,
%375025 Yerevan,
375049 Yerevan,
Armenia}
\author{A.~Ho}
\affiliation{
Department of Physics, San Diego State University,
5500 Campanile Drive, San Diego, California 92182, USA}
\author{E.E.~Kolomeitsev}
\affiliation{
School of Physics and Astronomy,
University of Minnesota,
Minneapolis,
MN 55455,
USA}
\author{M.C.~Miller}
\affiliation{
Department of Astronomy,
University of Maryland,
College Park,
MD 20742-2421, 
USA
}
\author{G.~R\"opke}
\affiliation{
Institut f\"ur Physik,
Universit\"at Rostock,
18051 Rostock,
Germany}
\author{J.~Tr\"umper}
\affiliation{
Max-Planck-Institut f\"ur extraterrestrische Physik,
85741 Garching, Germany}
\author{D.N.~Voskresensky}
\affiliation{
Gesellschaft f\"ur Schwerionenforschung mbH (GSI),
64291 Darmstadt,
Germany}
\affiliation{
Moscow Engineering Physical Institute,
Kashirskoe Shosse 31,%\\
11549 Moscow,
Russia
}
\author{F.~Weber}
\affiliation{
Department of Physics, San Diego State University,
5500 Campanile Drive, San Diego, California 92182, USA}
\author{H.H.~Wolter}
\affiliation{
Department f\"ur Physik,
Universit\"at M\"unchen,
85748 Garching,
Germany}

\begin{abstract}
A new scheme for testing nuclear matter equations of state (EsoS)
at high densities using constraints from neutron star (NS)
phenomenology and a flow data analysis of heavy-ion collisions
is suggested.
An acceptable EoS shall not allow the direct Urca process to occur in
NSs with masses below $1.5~M_{\odot}$,
and also shall not contradict flow and kaon production data of heavy-ion collisions.
Compact star constraints include the mass measurements
of $2.1\pm 0.2 ~M_{\odot}$ ($1\sigma$ level) for PSR J0751+1807 and of
$2.0\pm 0.1 ~M_{\odot}$ from the innermost stable circular orbit for 4U 1636-536,
the baryon mass - gravitational mass relationships from
Pulsar B in J0737-3039 and the mass-radius relationships from
quasiperiodic brightness oscillations in 4U 0614+09 and from
the thermal emission of RX J1856-3754.
This scheme is applied to a set of relativistic EsoS constrained
otherwise from nuclear matter saturation properties with the result
that  no EoS can satisfy all constraints, but those
with density-dependent masses and coupling constants appear most promising.
\end{abstract}

\maketitle

\section{Introduction}
The investigation of constraints for the high-density behavior
of nuclear matter (NM) %at supersaturation densities
has recently received new impetus when the plans to construct a
new accelerator facility (FAIR) at GSI Darmstadt were published.
Among others a dedicated experiment for the investigation of
the phase transition from hadronic matter to the quark-gluon plasma (QGP)
in compressed baryon matter (CBM) shall be hosted, which will
study the phenomena of chiral symmetry restoration and
quark (gluon) deconfinement accompanying the transition to the QGP.
A firm theoretical prediction for the critical baryon densities and
temperatures of this transition in the QCD phase diagram as well
as the existence and the position of a critical point depends sensitively
on both the properties of NM at high densities and
the model descriptions of quark-gluon matter in the nonperturbative
regime close to the hadronization transition.

In the present work we apply recently discovered astrophysical
bounds on the high-density behavior of NM in $\beta$- equilibrium,
i.e. neutron star matter (NSM), from compact star cooling
phenomenology and neutron star mass measurements together with information
about the elliptical flow in heavy-ion collisions (HICs) in order to
suggest a scheme for testing NM models.
This new test scheme will be applied to
candidates for the NM EoS which %are required to
describe properties at the saturation density
$n_s\approx 0.14 -  0.18~{\rm fm}^{-3}$
such as the binding energy per particle in
symmetric nuclear matter (SNM) $a_v$, the compressibility
$K$ and the asymmetry energy $J$
%together with their density dependent parts ($K'$ and $L$) constrained from
and characteristics of large nuclei, such as the neutron skin,
surface thickness
and spin-orbit splitting probing the domain of subsaturation densities.
In this paper we do not discuss the
possibilities of various phase transitions, like hyperonization,
pion and kaon condensations, quark  matter etc. 
\cite{Glendenning:1997wn,Weber:1999qn,Blaschke:2006xt}.
Corresponding comments on how their inclusion could affect our
results are  added at the appropriate places.

While there are several NM models giving a rather similar description of the
saturation and
subsaturation behavior they differ considerably in their extrapolations to
densities above $\sim 2 ~n_s$, the regime which is relevant for
NS physics and heavy-ion collisions.
Recent progress in astrophysical observations and new insights into the
compact star cooling phenomenology allow us to suggest in this paper a test
scheme for the high density EoS which consists of five elements.

The first one demands that any reliable nuclear EoS
should be able to reproduce the recently reported high pulsar mass
of $2.1\pm0.2 ~{\rm M_\odot}$ for PSR J0751+1807,
a millisecond pulsar in a binary system with a helium
white dwarf secondary \cite{NiSp05}.
Extending this value even to 2$\sigma$ confidence level
($^{+0.4}_{-0.5}~M_\odot$) means
that masses of at least $1.6~{M_\odot}$ have to be allowed.
Thus the EoS should be rather stiff to satisfy this constraint.

The second constraint has recently been suggested  in Ref.~\cite{Pod05} and
concerns pulsar~B in the
double pulsar system J0737--3039 which has the lowest
reliably measured mass for any NS to date, namely
$M=1.249 \pm 0.001 ~M_\odot$ \cite{Kramer:2005ez}.
If this star originates from the collapse of an ONeMg white dwarf
\cite{Pod05} and the loss of matter during
the formation of the NS is negligible, the baryon number, 
or equivalently the corresponding free baryon mass for the NS, 
has been determined to
$ 1.366 ~M_\odot \leq M_N \leq  1.375 ~M_\odot$.
It turns out that this constraint requires a rather strong binding of the
compact star.
A possible baryon loss of up to $1\%$ of $M_\odot$ during the formation of
the compact star broadens the corresponding baryon mass interval to
$ 1.356 ~M_\odot \leq M_N \leq  1.375 ~M_\odot$.

The next constraint emerges from recent results of NS cooling calculations
\cite{BlGrVo04}
and population synthesis models for young, nearby NSs \cite{Popov:2004ey}.
Following the arguments in Refs.~\cite{BlGrVo04,KoVo05}, direct Urca (DU)
processes, e.g., the neutron $\beta$-decay $n\to p+e^-+\bar\nu_e$,
produce neutrinos very efficiently.
The neutrino emissivities for these processes even with inclusion of nucleon
superfluidity effects are large enough that their
occurrence would lead to an unacceptably fast cooling of NSs
in disagreement with modern observational soft X-ray data in the
temperature - age  diagram.
According to these recent analyses, the DU process shall not occur in typical
NSs which have masses in the range of
$M_{typ} \sim 1.0 \div 1.5 ~M_\odot$, obtained from population syntheses
scenarios, see \cite{Popov:2004ey} and references therein.
This constrains the density dependence of the nuclear asymmetry energy which
should not be too strong.

The fourth constraint defines
an upper bound in the mass-radius plane for NSs,
derived from quasiperiodic oscillations (QPOs)
at high frequencies
of the low-mass X-ray binary (LMXB) 4U 0614+09 \cite{Miller:2003wa}.
For some LMXBs there is evidence for the innermost
stable circular orbit, which if confirmed suggests that the
masses of the NSs in many of these systems is between
$1.8\,M_\odot$ and $2.1\,M_\odot$ \cite{Barret:2005wd,Barret:2005zc}.

The fifth constraint comes from  a recent analysis of the thermal radiation
of the isolated pulsar RX J1856 which determines a lower bound
for its mass-radius relation that implies a
rather stiff EoS~\cite{Trumper:2003we}.

Finally, we include into the scheme constraints that are derived
from analyses of elliptic flow data and from kaon production in
heavy ion collisions.
Nuclear collisions have been described within a kinetic theory approach
and the results have been compared to experimental data for
the nucleon flow for densities up to $4.5\times n_s$ \cite{DaLaLy02}.
From this a region in the pressure-density diagram for SNM
has been given which defines upper (UB) and lower (LB) bounds to the
high density EoS and which is in accordance
with measurements of the elliptic flow.
Even though $4.5\times n_s$ is below typical central densities
that correspond to maximum masses of
NS configurations we use the fact that the
existence of such a region rules out rather stiff and very soft EsoS.

The outline of this work is the following.
In section \ref{sec:hadeos} we describe a set of modern
relativistic nuclear EsoS obtained within different approaches.
In the section \ref{sec:constraints}
the test scheme sketched above will be discussed in detail.
This includes the astrophysical constraints from the determination of
(maximum) NS masses in \ref{subsec:maxmassconstr},  the new
mass-baryon number test in \ref{subsec:M_N},
constraints for DU-cooling in \ref{subsec:du_constr} and
for the mass-radius relations of LMXBs in \ref{subsec:mr_qpo}
as well as the  mass-radius relation
from thermal emission of the isolated
NS RX J1856 in \ref{subsec:trumper}.
The EoS for SNM at supernuclear densities is constrained by  HIC experiments
from flow data analysis in \ref{subsec:flowconstr}, and  kaon production in
\ref{subsec:kaon}.
%and isospin diffusion in \ref{subsec:symen}.
In Section \ref{sec:cons} we derive two immediate consequences of this scheme:
a conjecture about a universal symmetry energy contribution to the EoS
in $\beta$-equilibrium and a sharpening of the flow constraint from HICs
using new information about the masses of compact stars.
A summary of the results of this work is given in section \ref{sec:firstconcl},
together with the conclusions to be drawn from them.

\section{Hadronic EoS}
\label{sec:hadeos}
\subsection{Model independent description}

There are numerous comparative studies of NM approaches for HIC and
NS physics applications in which a representation of the NM EoS has
been employed which is based on the nucleonic part of the binding energy
per particle given in the form
\begin{equation}
\label{eq:BA}
E(n,\beta)=E_0(n) + \beta^2 E_S(n),
\end{equation}
where $\beta = 1 - 2 x$ is the
asymmetry parameter depending on the proton fraction $x=n_p/n$
with the total baryon density $n=n_n+n_p$.
In Eq.~(\ref{eq:BA}) the function $E_0(n)$ is the binding energy in SNM,
and $E_S(n)$ is the (a)symmetry energy,
i.e. the energy difference between pure neutron matter and SNM.
Both contributions $E_0(n)$ and $E_S(n)$ are
easily extracted from a given EoS for the cases $\beta=0$ and $\beta=1$,
respectively.
The parabolic interpolation has been widely used in the literature, see e.g.
\cite{Lattimer:2000nx}.
It proves to be
an excellent parameterization of the asymmetry dependence for the purpose of
the present study and we will not go beyond it here.
{Nevertheless, it should be mentioned in this context that an 
{\em exact reproduction} of  a given EoS might require
higher order terms than $\beta^2$ which have been neglected here.}
From Eq.~(\ref{eq:BA}) all zero temperature EsoS of NM can be
derived by applying simple thermodynamic identities \cite{BaBu00}.
%Clipped and precise one has to determine
In particular, we obtain
\begin{eqnarray}
\varepsilon_B(n,\beta)&=&nE(n,\beta), \\
P_B(n,\beta)&=&n^2\frac{\partial}{\partial n}
E(n,\beta),\\
\label{eq:munp}
\mu_{n,p}(n,\beta)&=&
\left(
1+n\frac{\partial}{\partial n}
\right) E_0(n) \nonumber\\
&-&
\left(
\beta^2\mp2\beta-\beta^2n\frac{\partial}{\partial n}
\right) E_S(n)
\end{eqnarray}
for the baryonic energy density $\varepsilon(n)$ and pressure $P(n)$
as well as the chemical potentials of neutron $\mu_{n}$
(upper sign) and proton $\mu_{p}$ (lower sign), respectively.

NSM has to fulfill the two  essential conditions of
$\beta$-equilibrium
\begin{eqnarray}
\label{eq:betaeq}
\mu_n = \mu_p + \mu_e = \mu_p + \mu_\mu~~,
\end{eqnarray}
and charge neutrality
\begin{eqnarray}
\label{eq:neutr}
n_p -n_e-n_\mu =0~,
\end{eqnarray}
where $\mu_e$ and $\mu_\mu$ are the electron and muon chemical potentials, conjugate
to the corresponding densities $n_e$ and $n_\mu$.
In this paper we do not consider phase transitions to a deconfined phase at $n>n_s$.
If a first order phase transition were allowed a mixed phase could arise in
some density interval, see  \cite{Glendenning:1992vb}.
In general, the local charge neutrality condition could be replaced by the
global one. However, due to the charge screening this density interval
is essentially narrowed \cite{Voskresensky:2002hu,Maruyama:2005tb}.
The effect of the mixed phase on the EoS is also minor.

Due to Eq.~(\ref{eq:betaeq}) the  chemical potentials for muons and
electrons are equal, $\mu_\mu = \mu_e$ so that muons  appear in
the system, once their chemical potential exceeds their mass.
The EoS for NSM is considered as an ideal mixture of
a baryonic and a leptonic part,
\begin{eqnarray}
\label{eq:nsm}
\varepsilon(n, \beta)&=&\varepsilon_B(n, \beta)
+\varepsilon_e(n, \beta)+\varepsilon_\mu(n, \beta)~,
\\
P(n,\beta)&=&P_B(n,\beta)+P_e(n,\beta)+P_\mu(n,\beta)~.
\end{eqnarray}

Under NS conditions one parameter %chemical potential
is sufficient for a complete description,
e.g. the baryochemical potential $\mu_b$ which is conjugate
to the conserved baryonic charge.
In $\beta$-equilibrated NSM and in SNM
it is simply equivalent to the neutron
chemical potential, $\mu_b = \mu_n$.
Applying Eq.~(\ref{eq:munp}) and Eq.~(\ref{eq:betaeq})
shows that the electron and muon chemical potential
can be written as an
explicit function of baryon density and asymmetry parameter,
\begin{equation}
\label{eq:mue}
\mu_e(n,\beta)=4\beta E_S(n).
\end{equation}
Both electrons and muons are described as a
massive, relativistic ideal Fermi gas.

With the above relations only one
degree of freedom, namely the baryon density,
remains in charge neutral and $\beta$-equilibrated NSM at zero temperature.
Within this comfortable description
actual properties of NM
depend on the behavior of $E_0(n)$ and $E_S(n)$ only.
Both can be deduced easily from
any EoS introduced in the following section.

\subsection{Equations of state applied in this paper}

A wide range of densities up to and above ten times
the saturation density of NM is explored
in the description of NSs and HICs.
It is obvious that relativistic effects are important under these conditions.
Consequently, we study only nuclear EsoS that originate from
relativistic descriptions of NM. There are a number of different
approaches.

Phenomenological models are based
on a relativistic mean-field (RMF) description of NM with nucleons
and mesons as degrees of freedom \cite{Wal74,Ser86,Rei89,Rin96}.
The mesons couple minimally to the nucleons.
The coupling strengths are adjusted to properties of NM or atomic
nuclei.
A scalar meson ($\sigma$) and a vector meson ($\omega$)  are treated as
classical fields generating scalar and vector interactions.
The isovector contribution is generally represented by a vector meson $\rho$.
In order to improve the description of experimental data, a medium dependence
of  the effective interaction has to be incorporated into the model.
In many applications of the RMF model, non-linear (NL) self-interactions of
the $\sigma$ meson were introduced with considerable success
\cite{Bog77,Bog81,Rei86,Rei88,Gmu91,Sha93,Sug94,Lal97,Tok98,Lon04}.
This approach was later extended to other meson fields \cite{Tok98}.
As an alternative, RMF models with density-dependent nucleon-meson couplings
were developed \cite{Lon04,Fuc95,Typ99,Hof01,Nik02,Lal05}.
They allow for a more flexible description of the medium dependence and
several parameterizations were introduced recently.
In our study we choose two versions of the NL models with self-couplings of
the $\sigma$ meson field that were used in the simulation of HICs
\cite{Gaitanos:2003zg}.
In the parameter set NL$\rho$ the isovector part of the interaction is
described, as usual, only by a $\rho$ meson.
The set NL$\rho\delta$ also includes a
scalar isovector meson $\delta$ that is usually neglected in RMF models
\cite{Liu02}.
It leads to an increased stiffness of the neutron matter EoS and
the symmetry energy at high densities.
These particular NL models were mainly constructed to explore qualitatively
this scalar-isovector contribution in the symmetry energy.
However, they have no non-linearity or density dependence in the isovector
sector and lead to very, perhaps too, stiff symmetry energies at high densities.
The density dependent RMF models are also represented here by two parameter
sets \cite{Typel:2005ba}.
They are obtained from a fit to properties of finite nuclei (binding energies,
radii, surface thickenesses, neutron skins and spin-orbit splittings).
The parameterization DD is the standard approach with constrained rational
functions for the density dependence of the isoscalar meson couplings and an
exponential function for the $\rho$ meson coupling \cite{Typ99}.
In the D${}^{3}$C model additional couplings of the isoscalar mesons to
derivatives of the nucleon field are introduced that lead to a momentum
dependence of the nucleon self-energies that is absent in conventional RMF
model \cite{Typel:2005ba}.

Finally, we present a new parameterization of the RMF model with
density-dependent couplings that is fitted to properties of finite nuclei
(binding energies, charge and diffraction radii, surface thicknesses,
neutron skin in ${}^{208}$Pb, spin-orbit splittings) as in Ref.\
\cite{Typel:2005ba} with an additional flow constraint
(see below) by fixing the pressure of SNM to $P = 50$~MeV~fm${}^{-3}$ at
a density of $n=0.48$~fm$^{-3}$.
The density dependence of the $\sigma$ and $\omega$ meson coupling functions
is written as
\begin{equation}
\Gamma_i(n) = a_i
\frac{1+b_i(x+d_i)^2}
{1+c_i(x+d_i)^2}
\Gamma_i(n_{\rm ref}),
\end{equation}
where for the $\rho$ meson a simple exponential law
\begin{equation}
\Gamma_\rho(n) = \Gamma_\rho(n_{\rm ref})\exp [-a_\rho(x-1) ]
\end{equation}
is assumed.
The coupling constants $\Gamma_i(n_{\rm ref})$ have been fixed at a
reference density $n_{\rm ref}$.
The density dependent couplings are functions of
the ratio
$x=n/n_{\rm ref}$
with the vector density $n$. The parameters of this parameterization
called DD-F are specified in Table \ref{tab:DDF}.
For a more detailed description of these type of models see Ref.\
\cite{Typel:2005ba}.
\begin{table}[tb]
\vspace{3mm}
\begin{center}
\begin{tabular}{|c|cccccc|}
 \hline
 Meson & $m_{i}$  & $\Gamma_{i}(n_{\rm ref})$ &
$a_{i}$ & $b_{i}$ & $c_{i}$ & $d_{i}$ \\
 $i$ & [MeV] & & & & &
\\ \hline
 $\sigma$ & $555$        & $11.024$ & $1.4867$ & $0.19560$ &
            $0.42817$   & $0.88233$ \\
 $\omega$ & $783$        & $13.575$ & $1.5449$ & $0.18381$ &
            $0.43969$   & $0.87070$ \\
 $\rho$   & $763$        &  $3.6450$ & $0.44793$ &       &       &  \\
 \hline
\end{tabular}
\end{center}
\caption{\label{tab:DDF}Parameters of the DD-F model as defined in
Ref.\ \cite{Typel:2005ba} with a reference density of $n_{\rm ref}=
0.1469$~fm${}^{3}$.}
\vspace{3mm}
\end{table}

More microscopic approaches start from a given free nucleon-nucleon
interaction that is fitted to nucleon-nucleon scattering data and deuteron
properties.
In these {\it ab initio} calculations based on many-body techniques one
derives the nuclear energy functional from first principles, i.e., treating
short-range and  many-body correlations explicitly.
A successful approach to the nuclear many-body problem is the Brueckner
hole-line expansion.
In the relativistic Dirac-Brueckner-Hartree-Fock (DBHF)
approach \cite{honnef} the nucleon  inside the medium is dressed by the
self-energy $\Sigma$ based on a T-matrix.
The in-medium T-matrix which is obtained from the Bethe-Salpeter equation
plays the role of an effective two-body interaction which contains all
short-range and many-body correlations in the ladder approximation.
Solving the Bethe-Salpeter equation the Pauli principle is respected
and intermediate scattering states are projected  out of the Fermi sea.
The summation of the antisymmetrized T-matrix interactions with the
occupied states inside the Fermi sphere
yields finally the self-energy in Hartree-Fock approximation.
This coupled set of equations constitutes
a self-consistency problem which has to
be solved  by iteration.
It is possible to extract the nucleon self-energies from DBHF calculations
which can be compared with the corresponding quantities in phenomenological
RMF models, but this is not completely unambiguous as discussed in
Ref.~\cite{DaFuFae05}.
Here, we use recent results of (asymmetric) NM calculations in the DBHF
approach with the relativistic Bonn A potential in the subtracted T-matrix
representation \cite{boelting99,DaFuFae04,DaFuFae05,deJong:1997hr}.

In order to bridge the gap between fully microscopic and more phenomenological
descriptions that can be applied more easily to various systems, it is often
useful to adjust the parameters of the latter model to results extracted from
the former method.
As an example of this approach, we use a nonlinear RMF model (KVR)
with  couplings and meson masses depending on the $\sigma$- meson field
\cite{KoVo05}.
The parameters were adjusted to describe the SNM and NSM EoS of the
Urbana-Argonne group \cite{AkPaRa98}  at densities below four times the
saturation density.
Additionally, we study also a slightly modified parameter set (KVOR) of this
RMF model that allows higher maximum NS masses.
KVR and KVOR models elaborate the fact that not only the nucleon but also the
meson masses should decrease with increasing NM density.
Being motivated by the Brown-Rho scaling assumption, see \cite{Brown:1991kk},
and the equivalence
theorem between different RMF schemes, these models use only one extra
parameter compared to the standard NL RMF model (NL model).

The nuclear EsoS of these various models can be characterized
by comparing the parameters in the approximation of the binding energy per
nucleon
\begin{equation}
 E = a_{V} + \frac{K}{18} \epsilon^{2} - \frac{K^{\prime}}{162} \epsilon^{3} +
\dots + \beta^{2} \left( J + \frac{L}{3} \epsilon + \dots \right)
\end{equation}
around saturation as a function of the density deviation $\epsilon = (n-n_s)/
n_s$ and the asymmetry $\beta$.
In this form the EoS is characterized at saturation by the binding energy
$a_{V}$, the incompressibility $K$ and its derivative,
the skewness parameter $K^{\prime}$ and by the symmetry energy $J$ and the
symmetry energy derivative or symmetry pressure $L$ for asymmetric NM.
In table \ref{tab:nucmat} these parameters are given for the models employed
in this study.
Additionally, we give the Dirac effective mass $m_{D}=m-\Sigma$ (at the Fermi
momentum) in units of the free nucleon mass $m$
depending on the scalar self-energy $\Sigma$ of the nucleon.

\begin{table}[tb]
\vspace{3mm}
\begin{center}
\begin{tabular}{|l|ccccccc|}
 \hline
  Model         & $n_s$ & $a_{V}$   & $K$     & $K^{\prime}$ & $J$
& $L$
 & $m_{D}$\\
                & [fm$^{-3}$]   & [MeV]     & [MeV]   & [MeV]        &
[MeV]  & [MeV]  & [$m$]\\
 \hline
 NL$\rho$       & $0.1459$      & $-16.062$ & $203.3$ & $ 576.5$     &
$30.8$ & $83.1$
 & $0.603$ \\
 NL$\rho\delta$ & $0.1459$      & $-16.062$ & $203.3$ & $ 576.5$     &
$31.0$ & $92.3$
 & $0.603$ \\
 DBHF           & $0.1810$      & $-16.150$ & $230.0$ & $ 507.9$     &
$34.4$ & $69.4$
 & $0.678$ \\
 DD             & $0.1487$      & $-16.021$ & $240.0$ & $-134.6$     &
$32.0$ & $56.0$
 & $0.565$ \\
 D${}^{3}$C     & $0.1510$      & $-15.981$ & $232.5$ & $-716.8$     &
$31.9$ & $59.3$
 & $0.541$ \\
 KVR            & $0.1600$      & $-15.800$ & $250.0$ & $ 528.8$     &
$28.8$ & $55.8$
 & $0.805$ \\
 KVOR           & $0.1600$      & $-16.000$ & $275.0$ & $ 422.8$     &
$32.9$ & $73.6$
 & $0.800$ \\
 DD-F           & $0.1469$      & $-16.024$ & $223.1$ & $ 757.8$     &
$31.6$ & $56.0$
 & $0.556$ \\
 \hline
\end{tabular}
\end{center}
\caption{\label{tab:nucmat}Parameters of NM at saturation for
various EsoS (see text).}
\vspace{3mm}
\end{table}

There are significant differences between the models.
The saturation density $n_s$ in phenomenological models fitted to describe
atomic nuclei (DD, D${}^{3}$C, DD-F) is in the range
$0.147 - 0.15$~fm$^{-3}$.
The models NL$\rho$, NL$\rho\delta$ aiming at a description of low-energy HIC
data use the still smaller value $n_s \simeq 0.146$ fm$^{-3}$.
In contrast to that, the ``ab-initio'' {approach} (DBHF) shows a
saturation density that is considerably larger ($0.181$ fm$^{-3}$).
Approximations of the Urbana-Argonne type EoS (KVR, KVOR) use $0.16$ fm$^{-3}$.
The binding energy per nucleon is very similar in all models.
The incompressibility $K$ spans a rather wide range from soft
($K \approx 200$~MeV) to rather stiff ($K \approx 275$~MeV).
A major difference is found for the derivative $K^{\prime}$ of the
incompressibility that is relevant for the densities above saturation.
Models with parameters that are fitted to properties of finite nuclei
(DD, D$^{3}$C) %, especially surface properties, prefer
lead to a negative value of
$K^{\prime}$  %leading to
with a rather stiff EoS at higher densities.
It is well known that the ratio of the surface tension to the surface
thickness is determined by the parameters $K$ and $K^{\prime}$
\cite{St81,Fa88}, however, the exact relation depends on the assumption
for the shape of the surface.
{In the microscopic DBHF approach and the phenomenological models}
NL$\rho$, NL$\rho\delta$, KVR, KVOR
the parameter $K^{\prime}$ is rather large and, correspondingly, the EoS of
symmetric matter is softer at high densities.
The DD-F model constructed here is an exception.
In this parametrization we wanted to satisfy simultaneously
the description of finite nuclei
and the flow constraint that requires a soft EoS at high densities leading
to a very large  $K^{\prime}$.
Correspondingly, the surface properties are not optimally well described by
the DD-F model with clear systematic trends (radii too small for light nuclei
and too large for heavy nuclei, too small surface thicknesses as compared to
experimental data).
We also remark that the parameters of the nonlinear models NL$\rho$ and
NL$\rho\delta$
are not representative for conventional NL models that are fitted to
properties of finite nuclei, e.g., \ NL3, for which one finds $K=271.5$~MeV,
$K^{\prime}=-203.0$~MeV, $J=37.4$~MeV, $L=100.9$~MeV and $m_{D} = 0.596~m$
\cite{Lal97,Typel:2005ba}.

The symmetry energy $J$ is very similar for all models  with the exception of
a slightly  larger value in the DBHF calculation.
Here one has, however, to keep in mind that this value is read off at a
correspondingly larger density.
At $n=0.16~{\rm fm}^{-3}$ DBHF gives a value of  $J=31.5$~MeV, which is in
good agreement with the empirical models and also with the variational
approach of \cite{AkPaRa98}.

In contrast, the derivatives $L$ of the symmetry energy of the various models
are spread over a large range.
This quantity is closely related to the stiffness of the symmetry energy at
high densities.
In order to describe the experimental neutron skin thicknesses in atomic
nuclei a small slope of the neutron matter EoS is required
\cite{Typ01,Fur02,Chen:2005hw}.
Models with $L<60$~MeV (DD, D$^{}3$C, KVR, DD-F) fulfill this requirement
by introducing  an effective density dependence of the $\rho$ meson coupling
to the nucleon which goes beyond conventional NL RMF models.
A too small value for $L$ on the other hand seems to be in conflict with data
from isospin diffusion in heavy-ion collisions \cite{Steiner:2005rd} so
that recently from a combination of these data the limits
$62$~MeV~$<L<107$~MeV
have been suggested, see \cite{Li:2005sr} and Refs. therein.
{\em{Only the models NL$\rho$, NL$\rho\delta$, DBHF and KVOR satisfy this
requirement.}}
However, as {\em our emphasis is on high density
constraints of the EoS} we will not elaborate further on this
interesting point here but remark that it deserves a proper treatment.

The Dirac effective mass $m_{D}$ of the nucleon that appears in the
relativistic dispersion relation of the nucleons also shows a large variation
in the comparison.
In order to describe the spin-orbit splitting in atomic nuclei, a small value,
typically below {or around} $0.6~ m$ is required.
Parameter sets with larger values (KVR, KVOR) might have a problem
in this respect with the construction of a proper spin-orbit potential.
Larger values of the effective Dirac nucleon mass are motivated by fitting
the single nucleon spectra in nuclei \cite{KhSa82} with a large Landau mass
$m_L^*\simeq 0.9 - 1.0 ~m$.
The works \cite{JHM} find $m_L^*\simeq 0.74 - 0.82 ~m$ from the analysis of
neutron scattering off lead nuclei. The latter values relate to
$m_D \simeq 0.7 - 0.8 ~m$ \cite{G}. For a recent discussion of the momentum
and isospin dependence of the in-medium nucleon mass, see e.g.
Ref.~\cite{DaFuFae05}.

\begin{figure}[htb]
\includegraphics[width=0.42\textwidth, angle=-90]{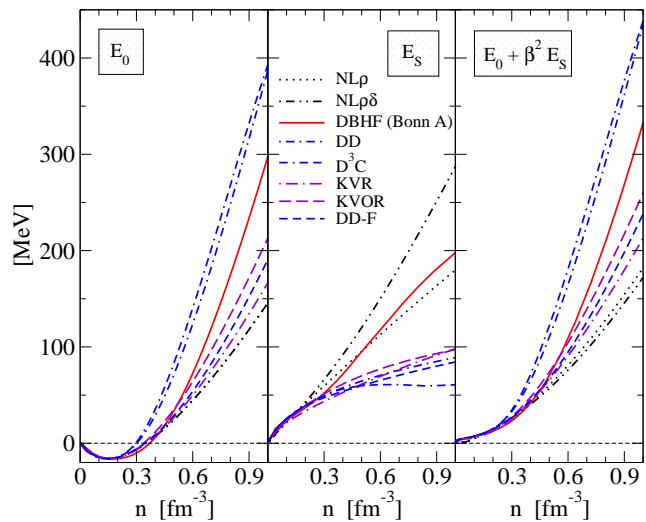}
\caption{\label{fig:e0_s0_nsm}
The energy per nucleon in SNM  $E_{0}(n)$ (left panel), the symmetry
energy $E_{S}(n)$ (middle panel) and the energy per nucleon in NSM
($\beta$-equilibrated and charge neutral) for the investigated models
(right panel). }
\end{figure}

The variation in the NM parameters is directly reflected in the behavior of
the energy per nucleon in SNM $E_{0}(n)$ and of the symmetry energy
$E_{S}(n)$ at densities above saturation as shown in
Fig.~\ref{fig:e0_s0_nsm}.
The various models of this study predict considerably different values for
$E_{0}(n)$ and $E_{S}(n)$ at high densities.
Under the condition of $\beta$-equilibrium, however,
the range of binding energy per nucleon
$E(n,\beta)$ shows a much smaller variation than expected from
$E_{0}(n)$ and  $E_{S}(n)$.
This is shown in the right panel of Fig.~\ref{fig:e0_s0_nsm} and discussed
further in Sect.~IV.

\section{Constraints on the EoS at high densities}
\label{sec:constraints}
In this section we will investigate to what extent the different EsoS
introduced in Sect.~\ref{sec:hadeos} fulfill the various constraints.
We postpone the discussion of the results of these tests to section 
\ref{sec:firstconcl} after two new consequences from our analysis are presented 
in Sect.~IV.

\subsection{Constraints from compact stars}

\subsubsection{Maximum mass constraint}
\label{subsec:maxmassconstr}
Measurements of ``extreme'' values,
like large masses or radii, huge luminosities etc.\
as provided by compact stars offer good opportunities to gain deeper insight
into the physics
of matter under extreme conditions as provided by compact stars.
Recent measurements on PSR J0751+1807 imply
a pulsar mass of $2.1\pm0.2\left(^{+0.4}_{-0.5}\right) {\rm M_\odot}$
(first error estimate with $1\sigma$ confidence, second in brackets with
 $2\sigma$ confidence) \cite{NiSp05}
which is remarkably heavy in comparison
to common values for binary radio pulsars
($M_{BRP}=1.35\pm 0.04~M_{\odot}$ \cite{Thorsett:1998uc}).
This special result constrains
NS masses to at least $1.6 ~M_{\odot}$ ($2\sigma$ confidence level)
or even $1.9 ~M_{\odot}$ within the $1\sigma$ confidence level.

The mass and structure of spherical, nonrotating stars,
to which we limit ourselves in this paper, is calculated
by solving the Tolman-Oppenheimer-Volkov(TOV)-equation,
which reads as
\begin{eqnarray}
\label{eq:TOV1}
\frac{{\rm d}P(r)}{{\rm d}r}
&=&
-~\frac{G[\varepsilon(r)+ P(r)]
[m(r)+ 4\pi r^3P(r)]}{r[r-2Gm(r)]}
\end{eqnarray}
where the gravitational mass $m(r)$ inside a sphere of radius $r$ is given by
\begin{eqnarray}
\label{eq:TOV2}
m(r)&=&
4\pi\int\limits_0^r{\rm d}r^\prime\;{r^\prime}^2\varepsilon(r^\prime)
\end{eqnarray}
which includes the effects of the gravitational binding energy.
The baryon number enclosed by that sphere is given by
\begin{eqnarray}
\label{eq:TOV3}
N(r)&=&
4\pi\int\limits_0^r \frac{{\rm d}r^\prime\;{r^\prime}^2 n(r^\prime)}
{\sqrt{1-\frac{2Gm(r^\prime)}{r^\prime}}} ~,
\end{eqnarray}
with $n(r)$ being the baryon density profile of the star.
Eq.~(\ref{eq:TOV1}) describes the gradient of the pressure $P$
and  implicitly the radial distribution
of the energy density $\varepsilon$ 
inside the star.
In order to solve this set of differential equations, one has to specify
the EoS, i.e., the relation between $P$ 
and $\varepsilon$ 
for which we take the EsoS introduced in the previous
Section \ref{sec:hadeos}.
We supplement our EsoS describing the NSs interior by an EoS for the crust.
For that we use a simple BPS model \cite{BaPe71}.
Due to uncertainties with different crust models one may obtain slightly
different mass-radius relations.

The stellar radius $R$ is defined by zero pressure at the stellar surface,
$P(R)=0$.
The star's cumulative gravitational mass is given then by  $M=m(R)$
and its total baryon number is $N=N(R)$.
In order to solve the TOV equations the radial change
of the pressure $P$ 
starting with  a given central value at radius $r=0$
has to be calculated
applying, e.g., an adaptive Runge-Kutta algorithm.

\begin{figure}[htb]
\includegraphics[width=0.42\textwidth,angle=-90]{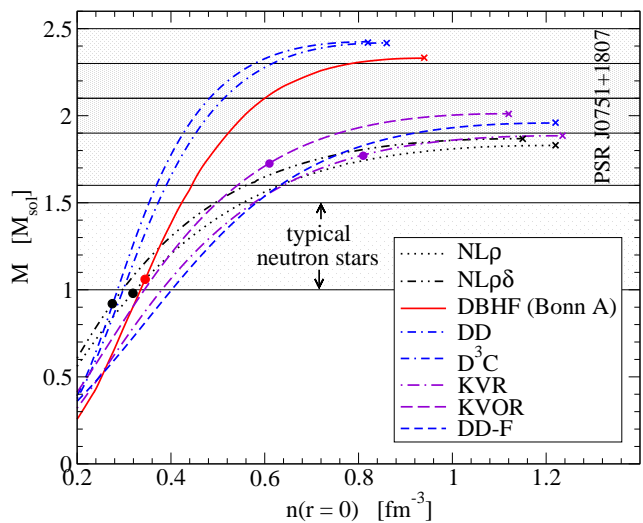}
\caption{\label{fig:m_n}
Mass versus central density for compact star configurations obtained by
solving the TOV equations (\ref{eq:TOV1}) and (\ref{eq:TOV2}) for all EsoS
introduced in Subsect. 2.2.
Crosses denote the maximum mass configurations, filled dots mark the
critical mass and central density values where the DU cooling process becomes
possible. According to the DU constraint, it should not occur in ``typical
NSs'' for which masses are expected from population synthesis
\cite{Popov:2004ey} to lie in the {lower} grey horizontal band.
The dark and light grey horizontal bands around 2.1 $M_\odot$ denote the
1$\sigma$ and 2$\sigma$ confidence levels, respectively, for the mass
measurement of PSR J0751+1807 \cite{NiSp05}.
 }
\end{figure}

The resulting NS masses as a function of their central density 
for the different EsoS are given in Fig.~\ref{fig:m_n} together with the 
mass range of typical NSs and the limits from PSR J0751+1807. 
Also shown in this figure are the points on the respective curves
where the DU process becomes possible, 
as further discussed in subsection~\ref{subsec:du_constr}.

The maxima of the mass-central density relations are easily determined then
and summarized in Table~\ref{tab:maxmass} 
for the EsoS investigated in this work.
As can be seen none of these values falls below the $2\sigma$ mass limit
of $1.6 ~M_\odot$, whereas the $1\sigma$ mass limit of $1.9 ~M_\odot$
would exclude NL$\rho$ and NL$\rho\delta$, while marginally excluding KVR.
%model {and marginally NL$\rho\delta$ and KVR}.
Thus the ability of this first and rather trivial test
to exclude a given EoS demands a high
accuracy of observations.
A more stringent test could be achieved
with  decreasing  error estimates
or the observation of at least one pulsar that is still more massive than
PSR J0751+1807.
{\em We point out that if a pulsar with a mass $M>2.1~M_\odot$ is
observed in the future, this will imply serious restrictions on the viable
EoS}, see Fig.~\ref{fig:m_n}.
Within the set of EsoS tested by us, only DD, D$^3$C and DBHF would survive.
Moreover, the maximum mass constraint is closely related
to the flow constraint.
This point will be further investigated within subsection
\ref{subsec:sharpflow}.

\begin{table}[htb]
\label{tab:mass}
\vspace{3mm}
\begin{center}
\begin{tabular}{|l|cc|c|c|c|c|}
\hline
\rule{0pt}{15mm}
{Model}
&
\turnt{$M_{\rm max}$  }
\turnt{$[M_\odot]$}
&
\turnt{$n_{\rm max}(0)$}
\turnt{$[{\rm fm}^{-3}]$}
&
\turnt{J0751+1807}
\turnt{(1$\sigma$) }
 &
\turnt{J0751+1807}
\turnt{(2$\sigma$)}
 &
\turnt{J0737-3039 B}
\turnt{(no loss)}
&
\turnt{J0737-3039 B}
\turnt{(loss 1$\% ~M_\odot$)}\\
\hline
NL$\rho$             & 1.83    &  1.22    & $-$  & $+$  & $-$ & $-$\\
NL$\rho\delta$       & 1.87    &  1.15    & $-$   & $+$  & $-$  & $-$     \\
DBHF            & 2.33         &  0.94    & $+$     & $+$  & $-$    & $+$    \\
DD              & 2.42         &  0.86    & $+$    & $+$  & $-$  & $-$       \\
D$^3$C          & 2.42         &  0.82    & $+$   & $+$  & $-$ & $-$ \\
KVR             & 1.89         &  1.24    & $\circ$  & $+$ & $-$ & $+$      \\
KVOR            & 2.01         &  1.12    & $+$  & $+$ & $-$ & $\circ$\\
DD-F            & 1.96         &  1.22    & $+$ & $+$ & $-$ & $+$ \\\hline
\end{tabular}
\end{center}
\caption{\label{tab:maxmass}
Maximum star masses, corresponding central densities and the fulfillment of
the strong ($1\sigma$) and weak ($2\sigma$) maximum mass constraint, as well
as the gravitational mass - baryon number constraint for Pulsar B in
J0737-3039 \cite{Pod05} without and with a mass loss of 0.01 $M_\odot$.
Fulfillment (violation) of a constraint is indicated with $+(-)$ and
a marginal result is rated with $\circ$.}
\vspace{3mm}
\end{table}

\subsubsection{Gravitational mass -- baryon number constraint}
\label{subsec:M_N}

\begin{figure}[htb]
\includegraphics[width=0.42\textwidth, angle=-90]{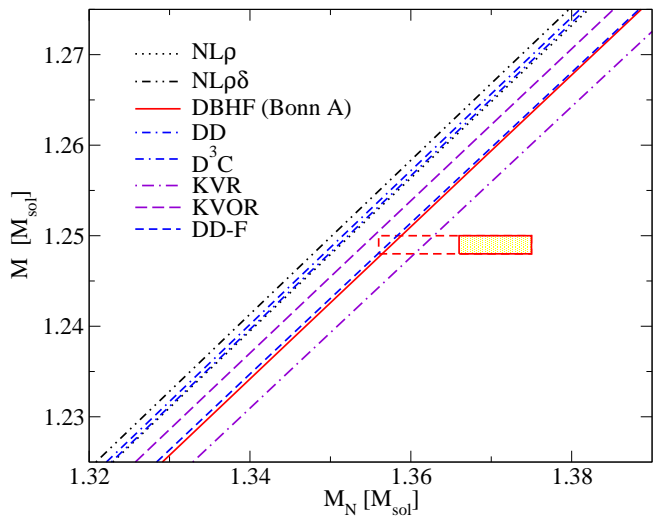}
\caption{
\label{fig:M_N}
Relation between gravitational mass $M$
and baryon mass $M_N$ (both in units of the solar  mass $M_\odot$)
of NSs for the EsoS discussed in this work.
The filled rectangle denotes the constraint derived
for Pulsar B in the double pulsar J0737-3039 \cite{Pod05}.
The empty rectangle demonstrates the change of the constraint when
the assumed loss of baryon number in the collapse amounts to 1$\%$
of the solar value.}
\end{figure}

Recently, it has been suggested in \cite{Pod05} that pulsar~B in the
double pulsar system J0737--3039 may serve to test models proposed
for the EoS of superdense nuclear matter. The system J0737--3039
consists of a 22.7~ms pulsars J0737--3039A (pulsar~A)
\cite{burgay03:a}, and a 2.77~ms pulsar companion J0737--3039B
(pulsar~B) \cite{lyne04:a}, orbiting the common center of mass in a
slightly eccentric orbit of 2.4 hours duration. One of the interesting
characteristics of this system is that the mass of pulsar~B is merely
$1.249 \pm 0.001 ~M_\odot$ \cite{Kramer:2005ez}, which is the lowest
reliably measured mass for any NS to date. Such a low mass
could be an indication that pulsar~B did not form in a type-II
supernova, triggered by a collapsing iron core, but in a type-I
supernova of an ONeMg white dwarf \cite{Pod05} driven hydrostatically
unstable by electron captures onto Mg and Ne.  The well-established
critical density at which the collapse of such stars sets in is $4.5\times
10^9~{\rm g}/{\rm cm}^3$ corresponding to an ONeMg core whose
critical baryon mass is  $M_N=N~u \sim 1.37 ~M_\odot$,
where the atomic mass unit $u=931.5$ MeV has been used \cite{Pod05}
to convert the baryon number to baryon mass.
Assuming that the loss of matter during the formation of
the NS is negligible, a predicted baryon mass for the
NS of $M_N = 1.366 - 1.375 ~M_\odot$ was derived in \cite{Pod05}.
This theoretically inferred baryon number range together
with the star's observed gravitational mass of $M = 1.249 \pm 0.001~M_\odot$
may represent a most valuable constraint on the EoS \cite{Pod05},
provided the above key assumption for the formation mechanism of the
pulsar B is correct.
Then any viable EoS proposed for NSM must predict a baryon number
in the range $1.366 \lsim M_N \lsim 1.375 ~M_\odot$ for a NS whose
gravitational mass is in the range $M=1.249 \pm 0.001~M_\odot$.
{\em None of the EsoS tested in this work satisfies this strong
constraint.}
The authors of \cite{Pod05} discussed caveats such as baryon loss and
variations of the critical mass due to carbon flashes during the collapse.
This constraint requires a very precise calculation of the baryon number,
e.g. a lowering of $M_N$ by 1$\%$ changes the outcome of this test
significantly.
Since the simulation of e-capture supernovae
and the evolution of their progenitors is still
a work in progress, more interesting results are expected in the near future.
The final value and accuracy of the baryon number of
J0737-3039 are therefore highly important.
The {result} of such calculations is shown in Fig.~\ref{fig:M_N} and
summarized in Table~\ref{tab:maxmass}.
{\em Finally we point out that this constraint is critically based
on the assumption of the formation scenario for pulsar B.
If this turns out to be incorrect the constraint has to be abandoned.}

\subsubsection{Direct Urca constraint}
\label{subsec:du_constr}

\begin{figure}[htb]
\includegraphics[width=0.42\textwidth,angle=-90]{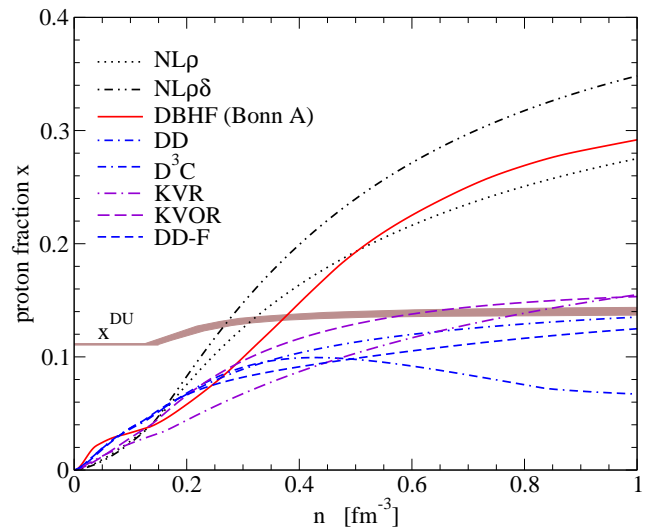}
\caption{ \label{fig:x_xDU}
Proton fractions $x=n_p/(n_n+n_p)$ for different EsoS.
The band labeled with $x_{DU}$
frames all threshold curves obtained for the investigated models.
According to Eq.~(\ref{eq:du_onset}) the range of possible threshold values
varies between  $11.1\%$ and $14.8\%$
{depending on the muon fraction}.
}
\end{figure}

The maximum mass constraint seems to have,
at least for the EsoS investigated in this paper,
a rather small exclusion potential.
The $M-M_N$ criterion, however, would 
provide more stringent limits only if 
the assumed formation mechanism of pulsar B in RX J0737-3039 and the
neglect of mass loss prove to be valid.
Adding the DU criterion will be seen to improve this scheme.

If the proton fraction $x=n_p/(n_p+n_n)$ exceeds a
critical value $x_{DU}$
the DU process $n\to p+e^-+\bar\nu_e$ becomes operative.
An estimate of this DU-threshold follows from
the triangle inequality for momentum conservation where
the moduli of the momenta are given by the neutron, proton and electron
Fermi momenta $p_{F_i}$. The typical neutrino energy 
of the order of the temperature $T$
is small and can be neglected.
In quasi-equilibrium $n\to p+e^-+\bar\nu_e$ implies that
$p_{F_n}\le p_{F_p} + p_{F_e}$.
From the charge neutrality condition $n_p = n_e + n_\mu$ one
easily finds the DU-threshold $x_{DU}$ as
\begin{equation}
\label{eq:du_onset}
x_{DU}=\frac{1}{1+(1+x_e^{1/3})^3},
\end{equation}
where $x_e=n_e/(n_e+n_\mu)$ is the leptonic electron fraction.
Since this depends on the symmetry energy, 
the DU-threshold is model dependent.
For $x_e=1$ (no muons) this formula reproduces the muon-free threshold value
of $11.1\%$~\cite{lattimer91}.
In the limit of massless muons, which is applicable for high densities
($x_e=1/2$) one finds an upper limit of $x_{DU} = 14.8\%$.
In Fig.~\ref{fig:x_xDU} the proton fraction 
as a function of density is shown for the different models, 
together with the DU-threshold value $x_{DU}$, 
given as a band for all the models.
As can be seen %from Fig.~\ref{fig:x_xDU}
this threshold can be reached for a wide
range of densities depending on the EoS.
For some models (DD, D$^3$C, DD-F) it does not take place at all.
The model-dependent DU-threshold
occurs at a
corresponding  critical baryon density.
Setting this as a star's central density
results in a DU-critical star mass $M_{DU}$.
These critical densities and DU-masses are marked in 
Fig.~\ref{fig:e0_s0_nsm} as  dots 
on those model curves, where the limit is reached.
Every star with a mass only slightly above $M_{DU}$
will be efficiently cooled  by DU-processes
and very quickly  becomes almost invisible for thermal detection
\cite{BlGrVo04}.
Nucleon superfluidity which suppresses the cooling rates has been included.
Values of the pairing gaps used in the literature have been used and then
varied to check the model dependence of the result  \cite{Grigorian:2005fn}.
Table \ref{tab:du} summarizes these DU critical masses for all models.
The DU constraint is fulfilled by the DD, D$^3$C and DD-F EoS models
which are not affected by the DU process at all and by KVR, KVOR which are
affected for masses higher than the limit for ``typical NS'' of $1.5~M_\odot$
obtained from population synthesis models \cite{Popov:2004ey}.
As a weaker constraint we also use $M_{DU}>1.35~M_\odot$.
This follows from both the population synthesis and the mass measurement of
binary radio pulsars.
If the DU process were allowed for $M_{DU}<1.35~M_\odot$ it would affect most
of the NS population. It should, however, not be expected that the objects
observed in X-rays were some exotic family of NSs rather than typical NSs.
DBHF and both NL models do not pass the DU test. They have a DU threshold
mass $M_{DU}<1.1~M_\odot$. Note also that only NSs with $M\gsim 1.1~M_\odot$
are produced in the standard scenario of NS formation in type II supernova
explosions \cite{Woosley}.

\subsubsection{Mass-Radius relation constraint from LMXBs}
\label{subsec:mr_qpo}
The kilohertz quasi-periodic brightness oscillations (QPOs)
seen from more than 25 NS X-ray binaries constrain
candidate high-density EsoS because there are
fundamental limits on how high-frequency such oscillations 
can be.  A pair of such QPOs is often seen from these systems
(see \cite{vdK:2000} for a general review of properties).  In
all currently viable models for these QPOs, the higher frequency
of the QPOs is close to the orbital frequency at some special
radius.  For such a QPO to last the required many cycles
(up to $\sim$100 in some sources), the orbit must obviously
be outside the star.  The orbit must also be outside the innermost
stable circular orbit (ISCO), because according to the predictions
of general relativity, inside the ISCO gas or particles would
spiral rapidly into the star, preventing the production of sharp
QPOs.  This implies \cite{Miller:2003wa,Miller:1996vj} that 
observation of a source whose maximum QPO frequency is $\nu_{\rm max}$
limits the stellar mass and radius to
\begin{equation}
\begin{array}{rl}
M&<2.2~M_\odot (1000~{\rm Hz}/\nu_{\rm max})(1+0.75j)\\
R&<19.5~{\rm km}(1000~{\rm Hz}/\nu_{\rm max})(1+0.2j)\; .\\
\end{array}
\end{equation}
Here $j\equiv cJ/GM^2$ (where $J$ is the stellar angular momentum)
is the dimensionless spin parameter, which is typically 0.1-0.2 for
these systems.  There is also a limit on the radius for any given mass.

These limits imply that for any given source, the observed $\nu_{\rm max}$
means that the mass and radius must fall inside an allowed ``wedge".
Therefore, any allowed EoS must have some portion of
its corresponding mass-radius curve fall inside this wedge.  The
wedge becomes smaller for higher $\nu_{\rm max}$, therefore the
highest frequency ever observed (1330~Hz, for 4U~0614+091; see
\cite{vanS:2000}) places the strongest of such constraints on the
EoS.  Note, though, that another NS could in principle have a greater 
mass and thus be outside this wedge, but an EoS
ruled out by one star is ruled out for all, since all NS have the same EoS. 
As can be seen from Fig. \ref{fig:M-R}, the current constraints from 
this argument do not rule out any of the EoS we consider.  
However, because higher frequencies
imply smaller wedges, future observation of a QPO with a frequency
$\sim 1500-1600$~Hz would rule out the stiffest of our EoS.  This
would therefore be a complementary restriction to those posed by
RX J1856.5-3754 (discussed below) and the implied high masses for some
specific NSs, which both argue against the softest EoS.

If one has evidence for a particular source that a given frequency
is actually close to the orbital frequency at the ISCO, then the
mass is known (modulo slight uncertainty about the spin parameter).
This was first claimed for 4U~1820--30 \cite{Zhang:1998}, but 
complexities in the source phenomenology have made this controversial.
More recently, careful analysis of {\it Rossi} X-ray Timing Explorer
data for 4U~1636--536 and other sources \cite{Barret:2005wd} has
suggested that sharp and reproducible changes in QPO properties are
related to the ISCO.  If so, this implies that several NSs
in low-mass X-ray binaries have gravitational masses between 
$1.9\,M_\odot$ and possibly $2.1\,M_\odot$ \cite{Barret:2005wd}. 
In Fig.~\ref{fig:M-R} we display the estimated mass $2.0\pm 0.1\,M_\odot$
for 4U~1636--536,
which would eliminate NL$\rho$ and NL$\rho\delta$ as
the softest proposed EoS even in the weak interpretation, and allow
only DBHF, DD and D$^3$C in the strong one, see Tab.~\ref{tab:du}.

\subsubsection{Mass-Radius relation constraint from RX J1856}
\label{subsec:trumper}
\begin{figure}[htb]
\includegraphics[width=0.42\textwidth, angle=-90]{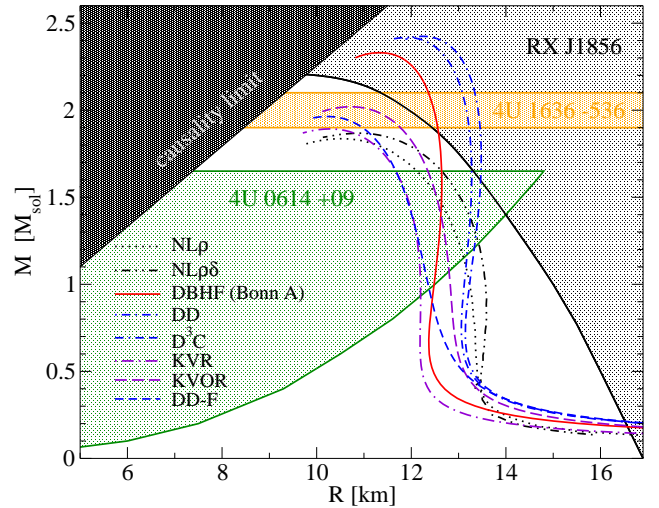}
\caption{
\label{fig:M-R}
Mass-Radius constraints from thermal radiation of the isolated NS 
RX J1856.5-3754 (grey hatched region) and from QPOs in the LMXBs 4U 0614+09 
(green hatched area) and 4U 1636-536 (orange hatched region) which shall be 
regarded as separate conditions to the EsoS.
For the mass of 4U 1636-536 a mass of $2.0\pm 0.1~M_\odot$ is obtained,
so that the weak QPO constraint would exclude the
NL$\rho$ and NL$\rho\delta$  EsoS whereas the strong one would leave only 
DBHF, DD and D$^3$C.
}
\end{figure}

After the discovery of the nearby isolated NS  RX J1856.5-3754 
(hereafter short: RX J1856) the analysis of its thermal radiation 
using the apparent blackbody
spectrum with a temperature $T_\infty=57$ eV
\cite{Pons:2001px} yielded a lower limit for the
photospheric radius $R_\infty$ of this object.
The distance of RX J1856 was initially estimated
to be 60 pc.
Since $R_\infty$ crucially depends
on this quantity a very small value of $ R_\infty\approx 8 $ km
was derived which could not have been explained
even with RX J1856 being a self-bound strange quark star \cite{Pons:2001px}.
The true stellar radius $R$ is given by $R_\infty = R(1-R/R_{S})^{-1/2}$,
with the Schwarzschild radius $R_S=2GM/R$.
New measurements predict a distance of at least $117$ pc,
which results in $R_\infty = 16.8$ km
and turns RX J1856 from the formerly smallest known NS
into the largest one \cite{Trumper:2003we}.
The resulting lower bound in the mass radius plane
is shown in Fig.~{\ref{fig:M-R}}.
There are three ways to interpret this result:
\begin{itemize}
\item[A)] RX J1856 belongs to compact stars with typical masses 
$M\sim 1.4 M_{\odot}$
and would thus have to have a radius exceeding
$14$ km (see Fig.~\ref{fig:m_n}).
{\em None of the examined EsoS can meet this requirement. }
\item[B)]  RX J1856 has a typical radius of $R \sim 12 - 13$ km, 
implying that the EoS has to be rather 
stiff at high density in order to allow for
configurations with masses above $\sim 2~M_\odot$.
In the present work this condition would be fulfilled for DBHF, DD and D$^3$C.
%These EoS, however, are in conflict with the DU cooling constraint.
This $M>1.6~M_\odot$ explanation implies that the object is very massive and
it is not a typical NS since most of NSs have $M<1.5~M_\odot$, as follows
from population synthesis models.
\item[C)] RX J1856 is an exotic object with
a small mass $\sim 0.2~ M_\odot$, which would be possible for all EsoS
considered here.
{\em No such object has been observed yet}, but some mechanisms for their
formation and properties have been discussed in the literature
\cite{Popov:2004nw}.

\end{itemize}
It cannot be excluded, however, that the distance measurement could be
revised by a future analysis. If the distance would turn out to be smaller
than the present value, then this constraint would have no discriminative
power any more since all EsoS could possibly fulfill it.
Should a revised distance value be larger than the present one, then only
the exotic low-mass star interpretation would remain which again is possible
for all (not self-bound) EsoS but which would raise the question about the
formation scenario for such a diffuse low-mass object.
Certainly this explanation of the puzzling object would no longer qualify
RX J1856 as an object to test the high-density nuclear EoS.
\begin{table}[htb]
\vspace{3mm}
\begin{center}
\begin{tabular}{|l|cc|cc|cc|ccc|}
\hline
\rule{0pt}{15mm}
Model
& \turnt{$M_{DU}$}
\turnt{$[M_\odot]$}
& \turnt{$n_{DU}(0)$}
\turnt{$[{\rm  fm}^{-3}]$}
& \turnt{$M_{DU}\ge 1.5~M_\odot$ }
& \turnt{$M_{DU}\ge 1.35~M_\odot$ }
& \turnt{4U 1636-536 (u)}
& \turnt{4U 1636-536 (l)}
& \turnt{RX J1856 (A)} %+ 4U0614}
& \turnt{RX J1856 (B)}
& \turnt{RX J1856 (C)}
\\
\hline
NL$\rho$        & 0.98          & 0.32    &$-$ &$-$  &$-$  &$-$     &$-$ 	&$-$ &$+$  \\
NL$\rho \delta$ & 0.92          & 0.28    &$-$ &$-$  &$-$  &$-$     &$-$ 	&$-$ &$+$  \\
DBHF            & 1.06          & 0.35    &$-$ &$-$  &$+$  &$+$     &$-$ 	&$+$ &$+$ \\
DD              & -             & -       &$+$ &$+$  &$+$  &$+$     &$-$& $+$ &$+$ \\
D$^3$C          & -             & -       &$+$ &$+$  &$+$  &$+$     &$-$ 	&$+$ &$+$ \\
KVR             & 1.77          & 0.81    &$+$ &$+$  &$-$  &$\circ$ &$-$ 	&$-$ &$+$ \\
KVOR            & 1.73          & 0.61    &$+$ &$+$  &$-$  &$+$     &$-$ 	&$-$ &$+$ \\
DD-F            & -             & -       &$+$ &$+$  &$-$  &$+$     &$-$ 	&$-$ &$+$  \\
\hline
\end{tabular}
\end{center}
\caption{\label{tab:du}
Critical compact star mass for the occurrence of the DU cooling process with
the corresponding central density, the criterion of the DU constraint
and the M(R) constraints from the isolated NS RX J1856.5-3754
and the low-mass X-ray binary 4U 1636-536 with its upper (u) and lower (l) 
mass limits, see text.
Symbols are defined in Table~\ref{tab:maxmass}.
}
\vspace{3mm}
\end{table}

Finally we want to emphasize another problem.
{Comparing Fig.~\ref{fig:M_N} and  Fig.~\ref{fig:M-R} we observe that models
producing a smaller radius (KVR, DD-F, DBHF) better accommodate the $M-M_N$
constraint and those having a larger radius (D$^3$C, DD) fulfill the
RX J1856 constraints better.
Out of the EsoS tested in this work only DBHF could satisfy these constraints
simultaneously and it would be a rather challenging task to resolve the problem of this 
EoS with the DU constraint. Further comments are given in the discussion below.}

\subsection{Constraints from heavy-ion collisions}

\subsubsection{The flow constraint}
\label{subsec:flowconstr}

The flow data analysis of dense SNM probed in HICs \cite{DaLaLy02}
reveals a correlation to the stiffness of the EoS which can be formulated as
a constraint to be fulfilled within the testing scheme introduced here.
\begin{figure}[htb]
\includegraphics[width=0.42\textwidth, angle=-90]{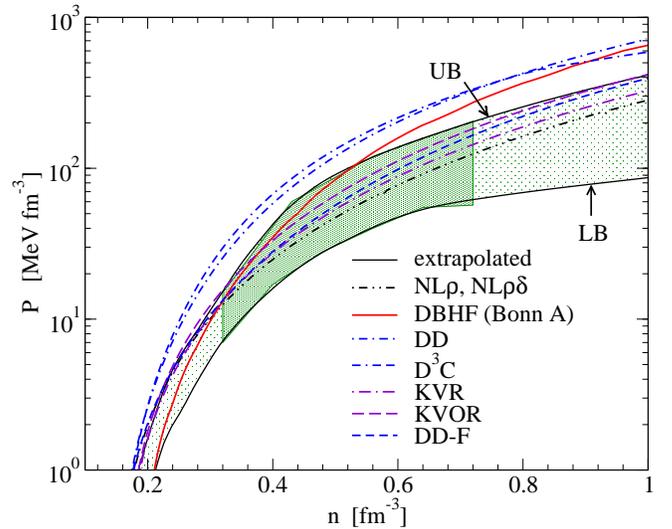}
\caption{
\label{fig:flow_extrapol}
Pressure region consistent with experimental flow data
in SNM (dark shaded region).
The light shaded region extrapolates this region to higher densities
within an upper (UB) and lower border (LB).
}
\end{figure}

The flow of matter in HICs
is directed both forward and perpendicular (transverse)
to the beam axis.
At high densities spectator nucleons may shield
the transversal flow into their direction and generate
an inhomogeneous density and thus a pressure profile
in the transversal plane.
This effect is commonly referred to as
elliptic flow and depends directly on the given EoS.
An analysis of these nucleon flow data, 
which depends essentially only on the isospin independent part of the EoS, 
was carried out in a particular model in Ref.~\cite{DaLaLy02}. 
In particular it was determined for which range of parameters 
of the EoS the model is still compatible with the flow data. 
The region thus determined is shown in  Fig.~{\ref{fig:flow_extrapol}} 
as the dark shaded region.
Ref.~\cite{DaLaLy02} then asserts that this region
limits the range of accessible pressure values at a given density.
For our purposes we extrapolated this region
by an upper (UB) and lower (LB) boundary,
enclosing the light shade region in Fig.~{\ref{fig:flow_extrapol}}.

Thus the area of allowed values does not represent experimental values itself,
but results from transport calculations
for the motion of nucleons in a collision \cite{DaLaLy02}.
Of course, it seems preferable to repeat these calculations
for each specific EoS,
but this would not be a manageable testing tool.
Therefore we adopt the results of ref.~\cite{DaLaLy02} as a
reasonable estimate of the preferable pressure-density domain in SNM.
Its upper boundary is expected to be stable against temperature
variations \cite{DaPrivat}.
The important fact is that the flow constraint
probes essentially only the symmetric part of 
the binding energy function $E_0(n)$.

Following Ref. \cite{DaLaLy02} the constraint arises for a density window 
between 2 and 4.5  times saturation density $n_s$. 
One has, however, to keep in mind that at high
densities this constraint is based on flow data from the AGS
energy regime ($E_{\rm lab} \sim 4 - 11$ AGeV). At these energies
a large amount of the initial bombarding energy is converted into
new degrees of freedom, i.e., excitations of higher lying
baryon resonances and mesons, which makes conclusions on the
nuclear EoS more ambiguous than at low energies. 
Nevertheless, the analysis of \cite{DaLaLy02} provides 
a guideline for the high density
regime which we believe to be reasonable.

As can be seen in Fig.~\ref{fig:flow_extrapol},
this last constraint is well fulfilled by the KVR, KVOR,
NL$\rho$ and NL$\rho\delta$ models.
For the latter two models this is rather obvious since they 
have already been tested
to reproduce flow data. %, as stated in section \ref{sec:hadeos}.
The constraint is satisfied for densities below $3~n_s$ by DBHF.
When comparing our models with the flow constraint below, we separate regions
of SIS ($n\lsim 3 n_s$) and AGS  ($n\gsim 3 n_s$) energies considering weak and
strong flow constraints.
DD and D$^3$C, which fulfilled the DU constraint well,
are significantly above the demanded region.
We want to emphasize that the DD-F model was constructed in this paper to pass
this test.
It is based on a reparametrization of the DD model,
in order to satisfy the introduced test scheme in most points.

\subsubsection{Constraints from subthreshold kaon production}
\label{subsec:kaon}

$K^+$ mesons were suggested as promising tools to probe the nuclear
EoS, almost 20 years ago \cite{aichelin85}.
The first channel to open in order
to produce a $K^+$  meson
is the reaction $NN\longrightarrow N\Lambda K^+$  which has a
threshold of $E_{\rm lab} =1.58$ GeV kinetic energy for the incident
nucleon.
When the incident energy per nucleon in a heavy ion reaction is below this
value one speaks of {\it subthreshold} kaon production.
This process is particularly interesting since it ensures
that the kaons originate from the high density phase of the reaction.
The missing energy has to be provided either by the Fermi motion of the
nucleons or by energy accumulating multi-step reactions.
Both processes exclude significant distortions from surface effects if one
goes sufficiently far below threshold.
In combination with the long mean free path subthreshold $K^+$
production is an ideal tool to probe compressed NM in relativistic HICs,
see Ref.~\cite{Fuchs:2005zg} for a recent review.

Within the last decade the KaoS Collaboration at GSI has performed systematic
measurements of the $ K^+$ production far below threshold
\cite{senger99,sturm01,kaos05}.
At subthreshold measurements which range from 0.6 to 1.5 AGeV laboratory
energy per nucleon compressions of  two to maximally three times $n_s$ are
reached.
Transport calculations have demonstrated that subthreshold  $K^+$
production provides a suitable
%and  reliable  %%% Shenya Kolomeitsev suggested to comment out
tool to constrain the EoS of SNM
at supersaturation densities \cite{Fuchs:2005zg,fuchs01,hartnack05}.
The theoretical analysis of the data implies a soft behavior of the EoS in
the considered density range
consistent with the flow constraint at moderate
densities ($n\lsim 3 n_s$) and supports DBHF, NL$\rho$, NL$\rho\delta$,
KVR, KVOR  and DD-F  EsoS in SNM \cite{hombach99,gaitanos01,stoicea04}.

\section{Consequences}
\label{sec:cons}
\subsection{Universal symmetry energy conjecture}

\begin{figure}[htb]
\includegraphics[width=0.42\textwidth, angle=-90]{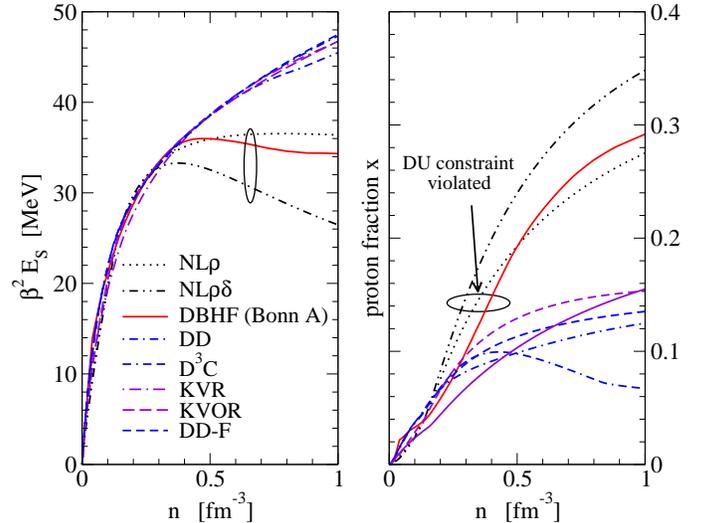}
\caption{
\label{fig:beta2S0_x_n}
Density dependence of the asymmetry contribution to the
energy per particle (left panel)
and of the proton fraction (right panel) in NSM.
Encircled curves correspond to EsoS that violate the DU-constraint.}
\end{figure}

Investigating the onset of
DU processes in section \ref{subsec:du_constr} has shown
that the DU threshold for the investigated models
can be reached for rather small baryon densities
slightly below $2n_s$ for NL$\rho$, NL$\rho\delta$, DBHF or, as the most
extreme opposite, not at all for DD, D$^3$C and DD-F.
Eq.~(\ref{eq:du_onset}) states that the threshold $x_{DU}$
depends on the electron-muon ratio.
The electron and muon densities are determined by
their chemical potentials.
In $\beta$-equilibrium $\mu_e=\mu_\mu$
is given in turn by Eq.~(\ref{eq:mue})
as a function of the asymmetry parameter $\beta$ and the symmetry energy $E_S$.
The resulting proton fraction $x$,
shown on the right hand side of Fig.~\ref{fig:beta2S0_x_n},
mainly maps the topological behavior of $E_S(n)$, see
Fig.~\ref{fig:e0_s0_nsm}.
As a rule of thumb therefore a large proton fraction $x$
is attained for stiff symmetry energies  $E_S(n)$.
The NL$\rho$, NL$\rho\delta$ and DBHF models confirm this rule
well and the symmetry energy used by these models can be sorted out for
contradicting the present cooling phenomenology,
as described in section \ref{subsec:du_constr}.
Fig.~\ref{fig:m_n} illustrates both DU- and maximum mass constraint.

The asymmetry contribution $\beta^2E_S(n)$ to the energy per nucleon
(left panel of Fig.~\ref{fig:beta2S0_x_n})
only shows a marginal dependence for different EsoS when compared
to differences in the energy per nucleon of SNM. % for the remaining model EoS.
These form a narrow band which allows two important statements.
First, the behavior of $\beta^2E_S(n)$ is to good approximation
universal for all EsoS which pass the DU-constraint.
The second conclusion regards the influence of the symmetry energy
on the mass of NS.
Here we find that due to the above universal behavior of $\beta^2E_S(n)$
the mass of a star is also dominated by the behavior of $E_0(n)$.
In other words it seems not very likely that it is possible to infer
any essential properties of $E_S(n)$
only from NS mass observations,
even if $E_0(n)$ would be perfectly known.
This point emphasizes the importance
of a more detailed understanding of the cooling behavior of compact stars
as an effective tool for probing $E_S(n)$ at high densities
beyond saturation.

\subsection{Sharpening the flow constraint}
\label{subsec:sharpflow}
As shown in sections \ref{subsec:maxmassconstr}
and \ref{subsec:flowconstr} the flow constraint
was more restrictive on the EoS 
than the maximum mass constraint
due to a large estimated error
for the mass of PSR J0751+1807.
But the flow constraint itself has
uncertainties, represented by the region
of possible pressures in Fig.~\ref{fig:flow_extrapol}.

\begin{figure}[htb]
\includegraphics[width=0.42\textwidth, angle=-90]{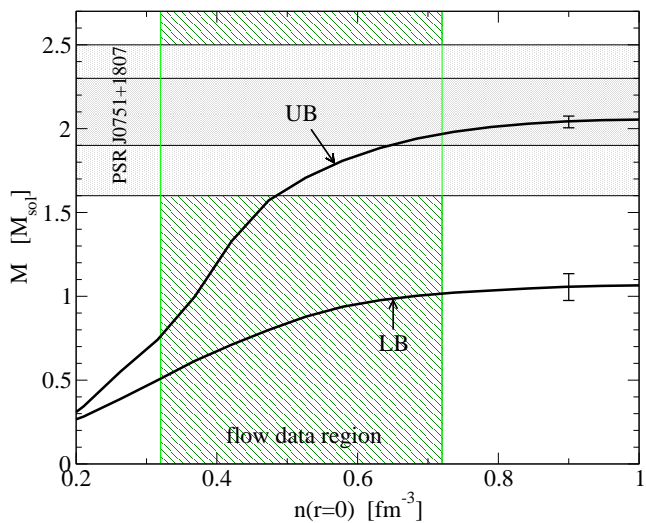}
\caption{\label{fig:flow_extrapol_mass}
{Mass versus central density for compact star configurations,
calculated using the UB and LB extrapolations of the flow constraint
boundaries from Fig.~\ref{fig:flow_extrapol}, $E_{0}(n)$,
together with different symmetry energies $E_{S}(n)$ not violating the
DU-constraint, see Fig.~\ref{fig:beta2S0_x_n}.}
The error bars illustrate the {maximum deviation resulting}
from choosing different symmetry energies $E_{S}(n)$.
The gray {horizontal} bars denote the expected mass of PSR J0751+1807
including for $1\sigma$ and $2\sigma$ confidence intervals, resp.,
whereas the vertical bars limit the density region
covered by the flow constraint.}
\end{figure}

Thus an EoS which is in accordance with
the flow constraint might still violate the maximum NS mass constraint.
This is demonstrated in Fig.~\ref{fig:flow_extrapol_mass},
where in particular the lower  boundary of
the limiting region in Fig.~\ref{fig:flow_extrapol}
was extrapolated to construct an artificial EoS (LB)
in order to obtain the mass-density relation for 
corresponding compact star configurations .

The resulting mass curve in Fig.~\ref{fig:flow_extrapol_mass}
is far from reaching even the weak mass constraint of $1.6 M_\odot$
originating from the lower $2\sigma$ bound on the mass of PSR J0751+1807.
It also cannot accommodate the well-measured mass
$M_{B1913+16}=1.4408 \pm 0.0003~M_\odot$ \cite{St04}.
This figure clearly demonstrates that the lower bound (LB) in
Fig.~\ref{fig:flow_extrapol}
does not satisfy the maximum mass constraint and
should be shifted upwards,
thus narrowing the band of the
flow constraint.

The maximum mass constraint demands
a certain stiffness of  $E_0(n)$  in order to obtain sufficiently large
maximum NS masses.
The small influence of $E_S(n)$ on the NS mass
can be well recognized on Fig.~\ref{fig:flow_extrapol_mass}, too.
Two different symmetry energies,
necessary to describe NSM,
were taken from the investigated EsoS.
They were chosen in accordance with
the DU-constraint and gave the largest (DD-F)
and smallest (D$^3$C) contribution to the binding energy at $n=1$ fm$^{-3}$.
The resulting deviations of the NS masses are shown 
as error bars on the curves in  Fig.~\ref{fig:flow_extrapol_mass}.
It results in a maximum difference of less than $0.2~M_\odot$ for
the mass curves corresponding to LB.
The same was done for an artificial EoS extrapolating the upper boundary (UB).
Here the largest error estimate of approximately $0.1~M_\odot$ is even smaller.
The maximum mass for UB of about $2.0~M_\odot$ again fulfills well
the corresponding constraint.

Comparing  the flow constraint in 
Fig.~\ref{fig:flow_extrapol} and 
the mass-radius constraint in Fig.~\ref{fig:M-R} we see that
none of the EsoS we tested satisfies both constraints.
Only DBHF is able to satisfy a weak flow constraint (for $n < 3 n_s$)
and the RX J1856 constraint under the assumption of an object with
$M>1.6~M_\odot$. Thus we again emphasize a problem for the RX J1856 constraint
with the joint fulfillment of other constraints as the $M-M_N$ and the
flow constraints.

\section{Summary and Conclusions}
\label{sec:firstconcl}
The task we intended with this work,
developing a test scheme for the nuclear EoS
by the present phenomenology of dense NM
in compact stars and heavy-ion collisions,
is satisfactorily completed at this point.
Applying this scheme to specific EsoS
offers some interesting insights
which indicate that astrophysical measurements
might become more important for the interpretation
of terrestrial measurements than presently accepted.
We have summarized the results of all suggested tests
performed on our choice of relativistic, high-density EsoS in
Table \ref{tab:sum} which reveals the discriminative power 
of their combined application in a broad
region of densities and isospin asymmetry.
We want to point out here, however, 
that each model was derived to describe a restricted region
in the $(n,\beta)$- plane and was not necessarily meant 
to describe a broader region.
In Tab. \ref{tab:sum} we rate the performance of the models when applied
nevertheless in a very wide $(n,\beta)$- region.
\begin{table}[htb]
\begin{center}
\begin{tabular}{|l||cc|cc|cc|cc|cc|cc||cc|}
\hline
{Model}                                 &
\turnt{$M_{\rm max}\ge 1.9~M_\odot$}            &
\turnt{$M_{\rm max}\ge 1.6~M_\odot$}            &
\turnt{$M_{DU}\ge 1.5~M_\odot$}         &
\turnt{$M_{DU}\ge 1.35~M_\odot$}        &
\turnt{4U 1636-536 (u)}& 
\turnt{4U 1636-536 (l)}& 
\turnt{RX J1856 (A)} & %+ 4U0614}               &
\turnt{RX J1856 (B)}                            &
\turnt{J0737 (no loss) }                &
\turnt{J0737 (loss 1\% $M_\odot$)}              &
\turnt{SIS+AGS flow constr.}                    &
\turnt{SIS flow+ K$^+$ constr.}                         &
\turnt{No. of passed tests}                 &
\turnt{(out of 6)}      \\
\hline
NL$\rho$ &$-$  &$+$  &$-$&$-$  &$-$  &$-$&$-$  &$-$  &$-$  &$-$  &$+$  &$+$ &$1$&$2$\\
NL$\rho\delta$ &$-$ &$+$ &$-$&$-$  &$-$ &$-$&$-$ &$-$ &$-$ &$-$ &$+$&$+$ &$1$&$2$\\
DBHF     &$+$  &$+$  &$-$&$-$  &$+$ &$+$ &$-$  &$+$  &$-$  &$+$  &$-$&$+$ &$2$&$5$\\
DD       &$+$  &$+$  &$+$ &$+$  &$+$ &$+$&$-$  &$+$  &$-$  &$-$  &$-$ &$-$  &$3$&$4$\\
D$^3$C   &$+$  &$+$  &$+$ &$+$  &$+$ &$+$ &$-$  &$+$  &$-$  &$-$  &$-$ &$-$  &$3$&$4$\\
KVR      &$\circ$ &$+$ &$+$ &$+$  &$-$ &$\circ$ &$-$  &$-$  &$-$  &$+$  &$+$ &$+$ &$3$&$5$\\
KVOR     &$+$  &$+$  &$+$ &$+$   &$-$ &$+$&$-$  &$-$  &$-$  &$\circ$&$+$&$+$&$3$&$5$\\
DD-F     &$+$  &$+$  &$+$ &$+$   &$-$ &$+$&$-$  &$-$  &$-$  &$+$  &$+$&$+$ &$3$&$5$\\
\hline
\end{tabular}
\end{center}
\caption{\label{tab:sum}
Summary of results for the suggested scheme of tests.
{Non separated columns show the results
for a strict (left) and weakened (right)
interpretation of the corresponding constraint.}
{The last column gives the total number of fulfilled
tests in the suggested scheme.} 
Symbols are defined in Table \ref{tab:maxmass}.
}
\end{table}

Due to its sensitivity to different contributions to the
energy this scheme motivates the necessity for changes in several EsoS if
one wants to apply them to the
whole available $(n,\beta)$
interval although they well describe properties at
the saturation density and a specific region of the $(n,\beta)$ plane.

In particular $E_S(n)$, the contribution of the symmetry energy
to the total energy, is probed by the DU-constraint.
It states that the DU process would cool
NSs much too fast, so that it
should occur for stars with masses greater than $\approx 1.35 - 1.5 M_\odot$
only.
If it would affect stars with masses below this limit
the DU-process would affect most of the known NS population.
We have shown that only EsoS with a rather soft symmetry energy at high density
fulfill the DU-constraint.

The symmetric matter contribution to the energy per nucleon $E_0(n)$ should
sufficiently describe the elliptic flow.
Adding the results of \cite{DaLaLy02}
to the test scheme, very soft EsoS are allowed.
The latter, however, can be sorted out by the maximum mass constraint.
As a result, these two combined constraints limit the stiffness
of a reliable EoS to be rather low (flow), but not too low at
$n\lsim 4 n_s$ and rather high for higher densities (maximum mass).

We want to stress that out of the tested EsoS only DBHF passes simultaneously 
the gravitational binding ($M-M_N$) as well as both mass-radius ($M-R$) tests.
The  $M-M_N$ constraint to be fulfilled requires a smaller radius of the
$M\simeq 1.25~M_\odot$ star, whereas the  $M-R$ test from RX J1856 favors 
substantially larger star radii, at least for $M\lsim 1.4~M_\odot$.
This contradictory situation would be resolved when RX J1856 is a star with a 
large mass or when an EoS would fulfill the  $M-M_N$ constraint and nevertheless
assign large radii to NS with typical masses. Such an EoS would be qualitatively 
different from the ones we investigate here.

The whole scheme left three of eight model EsoS, namely DD-F, KVR and KVOR
as most effective within a broad $(n,\beta)$ region under consideration.
The DD-F model explicitly fits properties of finite nuclei, especially
the neutron skin thickness of ${}^{208}$Pb that implies a small value of $L$.
The KVR model yields a similar value of $L$.
It might, however, have problems with respect to isospin diffusion
in heavy ion collisions since this small $L$ does not fit the
constraint deduced in Ref. \cite{Chen:2005ti} and \cite{Li:2005sr}.
The KVOR model fulfills this latter constraint. 
Here we point out that both KVR and KVOR were not applied to finite nuclei.
Thus it would be a challenge to apply such models to finite nuclei in the future.
All the models except DD-F demonstrate their predictive power within a broad
$(n,\beta)$ region whereas the DD-F model (as a modified DD model) has been
constructed in the present work in order to fulfill the flow constraint in
addition to constraints from saturation and finite nuclei common to DD models.
In contrast to
the phenomenological RMF models, DBHF is an ab initio
approach without room left for the readjustment of  free parameters.
But correlations beyond the ladder approximation are not
taken into account. In a certain sense, they are included,
although hidden in the fitted
parameters, in the phenomenological approaches.
However, an interesting aspect would be
to perform calculations for different types of free space
nucleon-nucleon interactions.
In particular the CD-Bonn potential  \cite{cdbonn}
which accounts more precisely for the isospin dependence of
the nuclear forces than Bonn A (used here) would be appropriate
for future investigations. Another point would be the explicit
inclusion of hyperonic degrees of freedom which may have
a significant impact on the NS matter EoS at high densities 
(depending on yet badly known nucleon-hyperon interaction).
This could open a
possibility for the DBHF and other EsoS to satisfy appropriately  the
DU constraint.

Beside the scheme's good overall selectivity
the joint application of different constraints
might give new interesting insights.
One of these is
the universal behavior of the contribution $\beta^2E_S(n)$
to the binding energy in NSM
we observed for all EsoS that fulfill the DU-constraint.
Then it seems to us that the flow constraint
limits the maximum mass of NSs to values around or not much about
the expected mass of PSR J0751+1807 with $M=2.1~M_\odot$, 
which also coincides with the upper mass limit for 4U 1636-536.
To verify this suggestion, a more detailed analysis,
similar to that shown in \cite{KaBa96}, has to be performed.

Next we want to emphasize that the maximum mass constraint
as a result of astrophysical measurements
further limits the pressure-density-region
which results from analysis of
elliptic flow data governed in terrestrial HIC experiments \cite{DaLaLy02}.
Although the introduced scheme would not change,
it seems useful to us to repeat these calculations under this point of view.
It would be interesting too,
to examine the agreement of experimental flow data
with numerically calculated values explicitly applying
the KVOR and DD-F models that have passed above constraints.

We have used here models which do not allow for phase transitions.
Any possible phase transition that may appear in the NSs interior
results in a decrease of the maximum NS mass.
All the models may well pass the constraint
$M_{\rm max}\gsim 1.6~M_\odot$ ($2\sigma$ uncertainty for PSR J0737-3039)
but KVR, and even DD-F and KVOR which successfully have passed most of  our
tests might get problems with the restriction
$M_{\rm max}\gsim 1.9~M_\odot$
($1\sigma$ level for PSR J0737-3039 and lower limit for 4U 1636-536) 
if the phase transition is sufficiently strong.

If the phase transition would occur in SNM it would also soften the EoS thus
modifying the flow constraint depicted in Fig.~4.
Nevertheless, if the energy gain due to the phase transition is not too large
the band of the flow constraint \cite{DaLaLy02} is rather broad and all the
models which already are situated within this band may still remain there.
However, if we considered the common ``flow+maximum mass'' constraint as
indicated by Fig.~6, the NL$\rho$ and NL$\rho\delta$ models could get a
problem crossing the lower boundary of the thus obtained new band, again if
the phase transition would be sufficiently strong.

The charged pion, kaon and the hyperonization transition in NSM change the
proton and electron concentrations thus affecting the DU threshold.
This threshold is then pushed up to higher densities.
Simultaneously, these transitions open new reaction channels, allowing for
DU reactions which involve new particles: $\pi^-$, $K^-$, hyperons and quarks.
The appearance of the condensates and /or filling of the new Fermi seas also
affects the values of the pairing gaps which are not well known.
With small gaps the new reaction channels lead to very rapid cooling of NS
raising the problem to appropriately describe the NS cooling.
However, if gaps are large these reactions might be non-operative and the DU
constraint may become softer or even non-effective.
The threshold densities for hyperonization strongly depend on poorly known
baryon-baryon interactions.
In case of repulsion the threshold density is pushed up \cite{KVk} and the
situation becomes more cumbersome.
We avoided studying such models here.

We postpone the analysis of EsoS allowing for phase transitions to future work.
Besides hyperonization, the possibility of a quark matter phase transition
should be studied.
This will be important for both
NSs and for the planned
HICs  in the planned CBM experiment at FAIR where large
baryon densities are to be created.

\section*{Acknowledgments}

The authors acknowledge discussions of the results presented in this work
at the workshops on ``The New Physics of Compact Stars'' at the ECT*
Trento, Italy, and on ``Dense hadronic matter and QCD phase transition'' in
Prerow, Germany.
T.K. has been supported in part by the
DFG Graduate School 567 on ``Strongly Correlated Many-particle Systems''
at Rostock University and by the Virtual Institute VH-VI-041 on
``Dense hadronic matter and QCD phase transition''of the Helmholtz
Association.  The work of D.N.V. has been
supported in part by DFG project No.  436 RUS 113/558/0-3
and T.G. from the BMBF grant No. 06ML981.
H.G. acknowledges funding from DFG grant No.  436 ARM 17/4/05,
E.v.D.  from grant FA 67/29-1.
E.E.K. was supported by the US Department of Energy under contract No.
DE-FG02-87ER40328 and the research of F.W. is supported by the National
Science Foundation (USA) under Grant PHY-0457329, and by the Research
Corporation (USA).  MCM was supported in part by a senior NRC fellowship
at Goddard Space Flight Center.

\end{document}